\documentstyle[12pt,epsf]{article}

\def\fnote#1#2{\begingroup\def\thefootnote{#1}\footnote{#2}\addtocounter
{footnote}{-1}\endgroup}
\newcommand{\AmS}{{\protect\the\textfont2
  A\kern-.1667em\lower.5ex\hbox{M}\kern-.125emS}}
\def\inbar{\vrule height1.5ex width.4pt depth0pt}
\def\IB{\relax{\rm I\kern-.18em B}}
\def\IC{\relax\,\hbox{$\inbar\kern-.3em{\rm C}$}}
\def\ID{\relax{\rm I\kern-.18em D}}
\def\IE{\relax{\rm I\kern-.18em E}}
\def\IF{\relax{\rm I\kern-.18em F}}
\def\IG{\relax\,\hbox{$\inbar\kern-.3em{\rm G}$}}
\def\IH{\relax{\rm I\kern-.18em H}}
\def\II{\relax{\rm I\kern-.18em I}}
\def\IK{\relax{\rm I\kern-.18em K}}
\def\IL{\relax{\rm I\kern-.18em L}}
\def\IM{\relax{\rm I\kern-.18em M}}
\def\IN{\relax{\rm I\kern-.18em N}}
\def\IO{\relax\,\hbox{$\inbar\kern-.3em{\rm O}$}}
\def\IP{\relax{\rm I\kern-.18em P}}
\def\IQ{\relax\,\hbox{$\inbar\kern-.3em{\rm Q}$}}
\def\IR{\relax{\rm I\kern-.18em R}}
\def\ZZ{\relax{\sf Z\kern-.4em Z}}
\def\fnote#1#2{\begingroup\def\thefootnote{#1}\footnote{#2}\addtocounter
{footnote}{-1}\endgroup}
\def\beq{\begin{equation}}
\def\eeq{\end{equation}}
\def\bea{\begin{eqnarray}}
\def\eea{\end{eqnarray}}
\def\lleq#1{\label{#1}\eeq}
 
\let\nn=\nonumber

\def\notin{\ \hbox{{$\in$}\kern-.51em\hbox{/}}}

\def\a{\alpha}        
     
\def\L{\Lambda}     
\def\Si{\Sigma}   \def\th{\theta}
 \def\cB{{\cal B}} \def\cC{{\cal C}}

 \def\cO{{\cal O}}

\def\lra{\longrightarrow}
\def\lolra{\longleftrightarrow}

\def\hbar{\bar h}

\def\ags{h^{(1,1)}}
\def\gens{h^{(2,1)}}
\def\kthree{K3}

 \def\II{{\bf II}}

\def\fth{{{\rm F}_{12}}}
\def\mth{{{\rm M}_{11}}}
\def\kthree{{\rm K3}}
\def\twoto{{{\rm T}^2}}
\def\fourto{{{\rm T}^4}}
\def\twoa{{\rm IIA}}
\def\het{{\rm Het}}
\def\cyfour{{{\rm CY}_4}}
\def\cythree{{{\rm CY}_3}}

\begin{document}
\hfill{\vbox{\hbox{hep-th/yymmnn}
             \hbox{UTTG-EP-97/9}
             \hbox{BONN--TH--97--3}
             \hbox{HUB--EP--97/8}}}

\vskip 0.9truein 
\centerline{\Large Dualities and Phase Transitions for }
\vskip .1truein
\centerline{\Large  Calabi-Yau Threefolds and 
       Fourfolds\fnote{\ddagger}{Based on a talk presented at the 
           30th Ahrenshoop International Symposium on the Theory of 
           Elementary Particles, Buckow, FRG, 1996}
       }

\vskip .5truein
 
\centerline{{\sc I. Brunner$^1$}\fnote{\star}{brunner@qft1.physik.hu-berlin.de}, 
        {\sc M.Lynker$^2$}\fnote{\diamond}{mlynker@siggy.iusb.edu}  
        and 
        {\sc R.Schimmrigk$^3$}\fnote{*}{Present address: Theory Group, 
           Department of Physics, University of Texas, 
         Austin, TX 78712}\fnote{\dagger}{netah@avzw02.physik.uni-bonn.de}
        }
       
\vskip .3truein
\centerline{\it $^1$Institut f\"ur Physik, Humboldt Universit\"at, 10115 Berlin, FRG}
\vskip .1truein
\centerline{\it $^2$Department of Physics and Astronomy, Indiana University}
\vskip .03truein
\centerline{\it  South Bend, 1700 Mishawaka Ave., South Bend, IN 46634, USA} 
\vskip .1truein
\centerline{\it $^3$Physikalisches Institut, Universit\"at Bonn, 53115 Bonn, FRG}

\vskip .9truein
\centerline{\bf Abstract}
\vskip .1truein

\noindent
We review several aspects of heterotic, type II, F-theory, and M-theory 
compactifications on Calabi-Yau threefolds and fourfolds. 
In the context of dualities we focus on the heterotic gauge structure 
determined by the various types of fibration relevant in the 
framework of heterotic/type II duality in D=4 as well as 4D F-theory. 
We also consider transitions between Calabi-Yau manifolds in both 
three and four dimensions and review some of the consequences for the 
behavior of the superpotential.

\renewcommand\thepage{}
\newpage

\baselineskip=18.2pt
\parskip=.1truein
\parindent=20pt
\pagenumbering{arabic}

\tableofcontents 

\vfill \eject

\section{Introduction}

In the last two years many dualities between perturbatively
different string theories have been discovered. An important
step was the discovery of a duality 
\beq
{\rm Het(T}^4) ~\lolra ~{\rm IIA(K3)}
\lleq{sixdual}
in six dimensions
between the heterotic string compactified on a four-torus
T$^4$ and the type IIA string compactified on a K3 surface 
\cite{ht94,w95,sen95,hast95}, extending an earlier observation 
of the equivalence of their moduli spaces \cite{six}.  
The non-abelian gauge structure of the heterotic string is related
to the ADE singularities of the K3 surface. 
A four-dimensional generalization of this duality  
\beq
{\rm Het(K3}\times {\rm T}^2) ~\lolra ~{\rm IIA(CY}_3) 
\lleq{fourdual} 
was formulated in \cite{kv95} by considering IIA on 
Calabi-Yau threefolds CY$_3$. Subsequently 
it was understood that the structure of the heterotic couplings 
requires the CY$_3$ on the type II side to be K3-fibrations 
$$
{\rm K3} ~\lra ~{\rm CY}_3 ~\lra ~\IP_1 
$$
with generic fibers K3 and base $\IP_1\sim S^2$ \cite{k3fib}. 
This suggests that to a certain extent the four-dimensional duality 
relation should be induced by a fiberwise application of the 6D 
duality \cite{vw95}. 
The detailed structure of the  heterotic gauge 
sector turns out to depend not only on the singularity 
structure of the fibers but also on the particular way these 
fibers are embedded into the threefold. The tool used in 
\cite{hs95} to determine the gauge group from the singularities 
and fibration structure is a bundle twist map, which constructs 
K3-fibered threefolds from a given K3 surface and an algebraic 
curve determined by the automorphism group of this surface. From 
the threefold point of view the twist map shows how 
the A$_{n}$ singularities of the 
K3 surfaces are tied together in the ADE Dynkin diagrams of 
the corresponding threefolds by singular curves on the 
threefold. We will review the salient features of this 
construction in Section 2 and describe some applications 
in Section 3.

Extending the net of dualities further by considering 
T-duality transformations between IIA and IIB theories as 
well as SL(2,$\ZZ$) duality in IIB, and taking into 
account the old relation between D$=$11 supergravity and IIA string 
theory in conjunction with the relation between IIA branes and
M-branes and Kaluza-Klein magnetic monopoles, it appears 
that many, if not all, dualities can be traced back 
to a few ancestors in 11D M-theory  \cite{ht94,w95,pt95,mth,rmth} 
and 12D F-theory \cite{cv96,mv96}.
Thus a picture arises in which all these different theories 
describe different regions of the moduli space of 
an underlying parent theory. 

M-theory is expected to be the eleven dimensional strong coupling 
limit of type IIA string theory. This has been emphasized in 
\cite{ht94,w95,pt95,mth,rmth} not only on the basis of the D$=$11 
Kaluza-Klein 
interpretation of the IIA effective field theory but also because 
the IIA brane structure follows from the electric M2-brane 
and its dual magnetic M5-brane induced by the 3-form of D$=$11 
supergravity. 
Further compactification of M-theory induces a number of known 
dualities. In six dimensions the relation (\ref{sixdual}) 
follows from the duality M(K3$\times S^1) \lolra $ Het(T$^4$) 
on the one hand and the relation of M-theory to IIA on the other.
We will describe more general compactifications of M-theory in 
Section 3. Some aspects of this unification of dualities have 
been reviewed in \cite{sen96}.

F-theory is a theory with a 12-dimensional origin which so far 
has been considered mostly through its compactifications to lower
dimensions.
The strong-weak coupling symmetry of type IIB theory was a motivation
to consider not only M-theory but also F-theory as a candidate
for a unifying theory. Recall that the SL(2,$\ZZ)$ symmetry of 
the type IIB string transforms the combination
 $\lambda = a+i~e^{-\phi}$ of the RR-scalar $a$ and the dilaton 
$\phi$ as the modular parameter of a torus. 
In the context of M-theory this symmetry  appears when  
M-theory is compactified on a torus and the modular parameter 
of this torus is compared with the coupling $\lambda $ of a 
type IIB string on a circle \cite{msl2z}.
F-theory provides us with a more geometric interpretation of 
this symmetry. The compactifications
of F-theory can be interpreted as special kinds of type IIB
compactifications where the coupling $\lambda $ is allowed to vary over
the internal manifold and to undergo SL(2,$\ZZ)$ monodromies.
More precisely, the internal manifold $X$ on which F-theory is
compactified, has to admit an elliptic fibration
\beq
{\rm T}^2~\lra~X~\lra~\cB.
\lleq{ellfibn}
The modular parameter of the fiber is identified with the coupling
of the IIB string, so that F-theory on $X$ is type IIB on $\cB $
with varying coupling constant. Because of the relation between
M-theory on T$^2$ and IIB on $S^1$ there is a duality between
F-theory on $X \times S^1$ and M-theory on X.

Consistent compactifications of M-theory and F-theory 
to three and four dimensions  respectively can be obtained 
via Calabi-Yau fourfolds CY$_4$. The resulting theories 
are N=1 supersymmetric theories. Again the fibration structure of
these fourfolds is important for possible duality relations in
lower dimensions. For F-theory in particular 
the general fibration structure  (\ref{ellfibn}) implies 
that the fourfold CY$_4$ has to admit an elliptic fibration.
If the base space $\cB$ in turn is a fibration of the form
$$
\IP _1 \lra \cB \lra S
$$
then type IIB on $\cB $ can be conjectured to be dual to a 
heterotic string on a Calabi-Yau threefold which is elliptically 
fibered over $S$ \cite{wsp96}.
 This conjecture is suggested by fiberwise application of the
duality between $\fth(\kthree)$  and the heterotic string on 
T$^2$ \cite{cv96}.  Altogether we then have the structure:
$$\matrix{{\rm T}^2 &\lra   &{\rm CY}_4 \cr
                    &       &\downarrow \cr
          \IP_1     &\lra   &\cB \cr
                    &       &\downarrow \cr 
                    &       &S, \cr 
 }$$
Here again the twist map is useful to construct fibrations from 
scratch by generalizing the constructions of \cite{hs95} to 
Calabi-Yau fourfolds \cite{bs96}. Particularly simple are 
CY$_3$-fibered fourfolds for which the fibers themselves are 
K3 fibrations of elliptically fibered K3 surfaces. The iterative 
fibrations of such manifolds show a 
nested structure which can be summarized in the diagram: 
\beq\label{nestfib}
  \matrix{{\rm T}^2 &\lra &{\rm K3} &\lra &{\rm CY}_3 
                              &\lra &{\rm CY}_4\cr
                  &     &\downarrow &   &\downarrow &     
                                 &\downarrow \cr
                  &     &\IP_1    &     &\IP_1      &     
                                      &\IP_1.  \cr}
\eeq
We will describe the twist map for Calabi-Yau $n$-fold 
fibered Calabi-Yau $(n+1)$-folds in Section 2. 
In Section 3 we apply the twist map to the discussion of 
heterotic/type II and F \& M theory duality relations involving 
the relation between the cohomology of Calabi-Yau threefolds 
\cite{hs95} and fourfolds and their dual heterotic gauge 
groups \cite{bs96}. 

Our final subject will be the unification of vacua and some 
applications to the generation of superpotentials in fourfold 
compactifications. 
Having been led to a new class of compactifications in terms of 
Calabi-Yau fourfolds an immediate question arises whether the 
recent understanding of the connectedness of string vacua generalizes
to the context of F-theory and M-theory. 
Much progress has been achieved recently in the interpretation 
of the earlier observation \cite{cdls88,ls95} that 
(weighted) projective complete intersection Calabi-Yau manifolds 
are connected via singular varieties involving only nodes. Such 
conifold transitions have now been understood physically  
in type II string theory via an effective field theory 
interpretation \cite{coniII} along the lines of Seiberg-Witten theory. 
 The connectedness of the space of Calabi-Yau threefolds 
   immediately implies the connectedness of certain regions of the 
  moduli space of Calabi-Yau fourfolds. For fibered fourfolds for which the generic 
   fiber is a Calabi-Yau threefold we can use the known conifold 
  transitions, or more severe transitions, to induce a transition in the 
 fourfold by degenerating the fibers.  Concrete ways to implement 
such fiber induced transitions have been described in \cite{bls96}.
 More general transitions are possible and in Section 4 we will review 
 \cite{bls96} how complete intersection fourfolds can be connected by transitions 
which are similar to the threefold conifold transitions of \cite{cdls88,ls95} 
but involve more severe degenerations of the varieties. 
 Similar to the conifold transitions of \cite{cdls88} the constructions 
 of \cite{bls96} are independent of any fibration structure of the fourfold. 
 As in the case of the threefold splitting transition which connects 
 fibered threefolds with nonfibered manifolds, the splitting transitions 
  between fourfolds also connects different types of Calabi-Yau spaces.  
In the final Section we show that it is possible in the process 
of such splitting transitions between fourfolds to generate 
nonvanishing superpotentials in M-theory compactifications 
\cite{bls96}. 

Several other aspects of M-theory and F-theory have been discussed 
recently in \cite{rmth} and \cite{morefth}.

\section{The twist map in arbitrary dimensions}

\subsection{The twist map}

In order to understand the way lower dimensional dualities can 
be inherited from the higher dimensional ones, and 
in particular to see to what extent this is possible at all, it 
is useful to have a tool which constructs  the necessary fibrations 
explicitly. One way to do this is by generalizing the orbifold 
construction of \cite{hs95} to arbitrary dimensions \cite{bs96}. 
In the following we 
will call this generalized 
map the twist map. Our starting point is a Calabi-Yau $n$-fold 
with an automorphism group $\ZZ_{\ell}$ whose action we denote 
by ${\bf m}_{\ell}$. Furthermore we choose
a curve $\cC_{\ell}$ of genus $g=(\ell -1)^2$ 
with projection $\pi_{\ell}: \cC_{\ell} \lra \IP_1$.  
The twist map then fibers Calabi-Yau $n$-folds into 
Calabi-Yau $(n+1)$-folds   
\beq
\cC_{\ell} \times {\rm CY}_n {\Big /}\ZZ_{\ell} \ni 
            \pi_{\ell}\times {\bf m}_{\ell}   
        ~~\lra ~~ {\rm CY}_{n+1}.
\eeq

For the class of weighted hypersurfaces 
\beq
\label{hypsubfibr}
\IP_{(k_0,k_1,...,k_{n+1})}[k]
~\ni ~\{y_0^{k/k_0}+p(y_1,...,y_{n+1})=0\}  
\eeq
for odd $k_0$ with  $\ell = k/k_0 \in \IN$ and $k=\sum_{i=0}^{n+1}k_i$,   
the cyclic action can be defined as 
\beq
\ZZ_{\ell}\ni {\bf m}_{\ell}:~ (y_0,y_1,...,y_{n+1})
~\lra ~(\a y_0,y_1,...,y_{n+1}), 
\eeq
where $\a$ is the $\ell^{th}$ root of unity. An algebraic 
representation of the curve $\cC_{\ell} $ is provided by 
\beq
\IP_{(2,1,1)}[2\ell] \ni \{x_0^{\ell} 
             -\left(x_1^{2\ell}+x_2^{2\ell}\right) =0\}
\eeq
with action $x_0 \mapsto \a x_0$ and the remaining coordinates 
are invariant.
The twist map in this weighted context takes the form     
\beq
\IP_{(2,1,1)}[2\ell] \times \IP_{(k_0,k_1,...,k_{n+1})}[k]
{\Big /}\ZZ_{\ell} ~\lra ~\IP_{(k_0,k_0,2k_1,...,2k_{n+1})}[2k] 
\lleq{twistn} 
and is defined as 
\beq
((x_0,x_1,x_2),(y_0,y_1,...,y_{n+1})) ~\lra ~
    \left(x_1 \sqrt{\frac{y_0}{x_0}},  
          x_2 \sqrt{\frac{y_0}{x_0}}, y_1,...,y_{n+1}\right).   
\lleq{twisthyp}
This map therefore embeds the orbifold of the product on 
the lhs into the weighted $(n+2)$-space as a hypersurface 
of degree $2k$. 
 
For $\ell=2$ the constructions of \cite{hs95, bs96} reduce 
to the Voisin-Borcea case \cite{cb97} 
and provide a more general class 
of fibrations for which the curve need not be the torus.

\subsection{Properties of the twist map}

The map (\ref{twisthyp}) shows that the structure of the 
$(n+1)$-fold is determined by a single $n$-dimensional fiber. 
This is of importance in the context of duality because it suggests that 
 lower dimensional dualities can be  
derived from known higher dimensional dualities \cite{vw95}. 
It indicates in particular that the degeneration structure 
of the fibers will play an important part in the determination 
of the gauge structure of the dual heterotic theories.

It is evident from (\ref{twisthyp})  however that the 
twist introduces new singularities on the  
$(n+1)$-fold fibration:  
the action of $\pi_{\ell}\times {\bf m}_{\ell}$
has fixed point sets which have to be resolved. This resolution
introduces new cohomology and therefore the heterotic gauge
structure is not completely determined by the $n$-fold fiber.

In the weighted category this aspect has two manifestations:
\begin{enumerate}
\item  If $k_0=1$, the action of $\ZZ_{\ell}$ generates a
    singular $(n-1)$-fold on the $n$-fold fiber. 
      This  $(n-1)$-fold singular set, which is not present in the 
     original $n$-fold, plays different r\^{o}les in  
      different dimensions.   
      In the case of K3-fibered threefolds  
    the effect of the resulting singular curve is 
      to introduce additional branchings in the
      resolution diagram of the Calabi-Yau threefold. It is this
      branching which determines the final gauge structure of the
      heterotic dual of the IIA theory on the threefold.
      
      For CY$_3$-fibered Calabi-Yau fourfolds such additional 
     branchings do not necessarily occur.  
\item More generally one encounters $k_0>1$, considered for
   threefolds in \cite{ls95} and for fourfolds in \cite{bs96,bls96}.
   In such a situation the orbifolding $\ZZ_{\ell}$ generates further
    singularities on the threefold and the fourfold.
    Depending on the structure of the weights these additional singularities 
   on $n$-folds can have dimensions $0,...,(n-1)$.
     Thus for threefolds we can have additional singular points or curves
and for fourfolds there can be points, curves or surfaces.
\end{enumerate}

If we start from a particular $n$-fold there are in general
several possibilities to pick an automorphism and the
corresponding curve and to construct a twist map out of this.
The resulting images of the twist maps will have different
singularities and therefore different cohomology. We will illustrate 
this for threefolds and fourfolds in the following Section.

\section{Applications of the twist map}
In this Section we will apply the twist map to derive the 
gauge structure of fibered Calabi-Yau threefolds and fourfolds 
 in the context of heterotic/type II duality and 4D F-theory 
respectively.

\subsection{4D heterotic/type II duality}

The dualities (\ref{sixdual}) and (\ref{fourdual}) are rather 
brief short hand notation for relations that involve somewhat 
more elaborate constructions. For 4D dualities what is meant 
by  Het(K3$\times $T$^2)$ in the case of the E$_8\times $E$_8$ 
string is a two step compactification in which the string is 
first compactified on a torus T$^2$. This leads to a grand 
gauge group of E$_8\times $E$_8\times G\times $U(1)$^2$ where 
$G$ is the group induced by the torus. Generically $G=$U(1)$^2$ 
but at special radii this symmetry becomes enhanced. The second 
step involves further compactification on K3 and the choice of a
vector bundle $\oplus_i V_i \lra $K3 such that anomaly 
cancellation holds $\sum_i c_2(V_i)=c_2$(TK3). Embedding the 
structure group into the gauge group provides the starting point 
for a whole cascade of models obtained by repeated 
Higgsing \cite{kv95}. 
 More precisely the relations then take the form
$$
{\rm Het}_{{\rm E}_8\times {\rm E}_8}(\oplus_i V_i 
 \rightarrow  {\rm K3}\times {\rm T}^2)_{\rm Higgsed} 
~~ \lolra ~~{\rm IIA(CY}_3). 
$$
We will describe some examples further below. 

The virtue of heterotic/type IIA duality is that it allows an 
exact computation of the vector multiplet couplings in the 
heterotic theory. This result is not obvious because the dilaton 
of the heterotic theory sits in a vector multiplet and therefore 
the vector multiplet moduli space receives corrections from 
spacetime instantons. Under the dualities (\ref{sixdual}) and 
(\ref{fourdual}) it is however identified with the vector multiplet 
moduli space of the type IIA string. In type IIA the dilaton 
sits in a hyper multiplet and therefore its vector multiplet 
does not receive corrections by spacetime instantons. Compactifying 
IIA on a Calabi-Yau threefold associates the vector multiplets 
with K\"ahler deformations living in H$^{(1,1)}$(CY$_3$) and 
therefore these couplings are corrected by worldsheet instantons. 
Via mirror symmetry we can however analyze the vector couplings 
of a type IIA(CY$_3$) theory by computing the exact hyper multiplet 
couplings of IIB compactified on the mirror of CY$_3$.
Thus 4D heterotic/type IIA duality in tandem with mirror symmetry 
makes it possible to determine nonperturbative heterotic corrections 
by doing a tree-level computation in the type II theory.

For type IIA string theory compactified on Calabi-Yau threefolds 
CY$_3$ we are interested in the twist map for $n=2$. Even though 
the possibilities are somewhat limited compared to higher 
dimensions because there is just one two-dimensional Calabi-Yau 
space, the CY$_2=$K3 surface, it is possible to construct a great 
many K3-fibered threefolds by choosing different realizations of 
this surface with their associated automorphisms. Starting from 
the heterotic/type IIA duality (\ref{sixdual}) in six dimensions 
we know that the heterotic gauge structure in D$=$4, determined 
by the vector multiplets in type IIA(CY$_3$) \cite{kv95},  
is parametrized by the second cohomology group H$^{(1,1)}$(CY$_3$) 
of the K3-fibered threefolds.   

\relax From (\ref{twistn}) with $n=2$ we see that this group is 
determined by a single K3 surface and the action of the 
automorphism $\ZZ_{\ell}$ on this surface and the curve 
$\cC_{\ell}$. In the simplest 
instances when the orbifold resolution of the resulting threefold 
does not introduce new cohomology off the fibers the  
gauge group is specified by the invariant part of the Picard 
lattice with respect to the action defining the fibration.

\noindent
{\bf Example I:}
We start with the K3 Fermat surface $K$ in the weighted 
configuration $\IP_{(1,6,14,21)}[42]$. $K$ has an automorphism 
group $\ZZ_{42}$ and the associated curve is 
$\cC_{42}=\IP_{(2,1,1)}[84]$.
The image of the twist map is $\IP_{(1,1,12,28,42)}[84]$.
The twist has introduced a $\ZZ_2$-singular curve 
$C=\IP_{(6,14,21)}[42]$.
On top of the curve $C$ one finds a $\ZZ_2$, a $\ZZ_3$
and a $\ZZ_7$ fixed point, leading to 1, 2 and 6 new (1,1)-forms,
respectively. Hence we have a total of $h^{1,1}=11$.
 Each of these resolutions leads to an  
$A_n$ resolution diagram with $n=1,2,6$ respectively, i.e.
\[\unitlength .5cm
\begin{picture}(11,7)
\put(4,1){\line(1,0){5}}
\put(4.5,3){\line(1,0){2}}
\put(5.5,4){\line(1,0){2}}
\put(9.5,4){\line(1,0){2}}

\put(5,3.5){\line(0,-1){3}}
\put(6,4.5){\line(0,-1){2}}
\put(7,5.5){\line(0,-1){2}}
\put(8.5,5.5){\line(0,-1){2}}
\put(10,5.5){\line(0,-1){2}}
\end{picture}
\]
This is the blow-up structure as it appears on the fiber. 
On the threefold 
these exceptional divisors all sit on the curve 
$C$ and therefore $C$ links up the $A_n$ branches into 
the tree diagram  
\[\unitlength .5cm
\begin{picture}(11,7)
\put(4,1){\line(1,0){5}}
\put(4.5,3){\line(1,0){2}}
\put(5.5,4){\line(1,0){2}}
\put(9.5,4){\line(1,0){2}}

\put(5,3.5){\line(0,-1){3}}
\put(6,4.5){\line(0,-1){2}}
\put(7,5.5){\line(0,-1){2}}
\put(8.5,5.5){\line(0,-1){2}}
\put(10,5.5){\line(0,-1){2}}\thicklines
\put(11.5,5.5){$C$}
\put(11.4,5.4){\vector(-3,-1){.8}}
\put(6.5,5){\line(1,0){4}}
\end{picture}
\]
The resolution diagram is given by Dynkin diagram of 
E$_8\times $U(1)$^2$. 
The heterotic dual of this model is an $E_8 \times E_8$
string compactified on K3 $\times $ T$^{2}$ by embedding all 24 instantons in
the first $E_8$ factor. This $E_8$-factor is completely
higgsed whereas the second $E_8$ remains unbroken.
The radii of the torus are not fixed at some
particular symmetric point.
Altogether, we obtain a gauge group E$_8 \times U(1)^4 $.
The heterotic spectrum contains 11 vectormultiplets plus the
graviphoton and 492 hypermultiplets, in agreement with the
Hodges $(\ags , \gens)$ of the threefold. Also the resolution
diagram agrees with the heterotic gauge group. It is entirely
determined by the Picard lattice \cite{k3fib}.

This manifold was also considered in the context of F-theory
compactifications to six dimensions. It is the element $n=12$
of the series $\IP_{(1,1,2,2n+4,3n+6)}[6(n+2)]$ which was
considered in \cite{mv96}.

\noindent 
{\bf Example II:} 
In our second example we will illustrate that the same K3-fiber 
can lead to different Calabi-Yau threefolds, depending on 
the twist map chosen to construct the fibration.
In general a single $\kthree $ surface can lead to many 
threefolds with different Hodge diamonds. 

Consider first the Fermat surface in the K3 configuration 
$\IP _{(1,1,1,3)}[6]$. The Fermat polynomial has an automorphism 
$\ZZ _6 : (y_0, y_1, y_2, y_3) \mapsto 
(\alpha y_0, y_1, y_2, y_3 ) $ where $\alpha $ is a $6^{th} $ root
of unity. The associated curve is  
$\cC _{6} = \IP _{(2,1,1)}[12] $.
Applying the twist map leads to the well-known 
two-parameter Calabi-Yau manifold in the threefold 
configuration $\IP _{(1,1,2,2,6)}[12]$ considered by 
Kachru and Vafa \cite{kv95}.
This manifold has a $\ZZ_2$-singular curve $\IP_{(1,1,3)}[6]$ whose
resolution leads to an additional $(1,1)$-form. There are no
further singularities and therefore this manifold 
has Hodge numbers $(h^{(1,1)}, h^{(2,1)})= (2,128)$.

 A different threefold can be obtained by starting again from 
the Fermat hypersurface in $\IP _{(3,1,1,1)}[6]$ but alternatively 
using the automorphism $\ZZ _{2} : (y_0, y_1, y_2, y_3) \mapsto
(\alpha y_0, y_1, y_2,  y_3 )$,  where $\alpha \in \{ 1,-1 \} $. The
corresponding curve is the torus $\cC _{2} = \IP _{(2,1,1)}[4]$. 
Using this curve
we obtain a different twist map, resulting in the manifold
$\IP _{(3,3,2,2,2)}[12] $ with Hodge numbers $(\ags ,\gens)
= (6,60)$, which  therefore is topologically distinct from the 
degree twelve hypersurface considered above. 
 Again the image of the twist map has a
singular $\ZZ _{2}$ curve, giving one  $(1,1)$-form.
But in addition there are four $\ZZ _{3} $ points $\IP_1[4]$ 
which do not lie on the singular curve.
The resolution of these additional singularities leads to
four additional $ (1,1)$-forms. 

The change in the Hodge diamond is even more pronounced in the case
of Calabi-Yau fourfolds in the context of M-theory and F-theory 
to which we will turn now.

\subsection{4D F-theory and 3D M-theory on Calabi-Yau fourfolds}

In the context of F-theory we are interested in pushing down 
the 8-dimensional duality \cite{cv96} 
\beq
{\rm F}_{12}{\rm (K3)}~~ \lolra ~~{\rm Het(T}^2)
\lleq{feight}
on elliptically fibered K3s to D$=$6 compactification 
\beq
{\rm F}_{12}{\rm (CY}_3)~~ \lolra ~~{\rm Het(K3)}
\lleq{fsix}
on elliptic threefolds and, finally, to D$=$4 compactifications 
\beq
{\rm F}_{12}{\rm (CY}_4)~~ \lolra ~~{\rm Het(CY}_3)  
\lleq{ffour}
on elliptic Calabi-Yau fourfolds. Compactifying these relations 
further  on a torus we obtain the diagram 
$$
\matrix{ \fth(\kthree \times \twoto)
                                     &\lolra &\mth(\kthree \times S^1)  
                                     &\lolra &\twoa(\kthree)
                                     &\lolra &\het(\fourto)  \cr
         \downarrow
                                    &        &\downarrow
                                    &        &\downarrow
                                    &        &\downarrow \cr
         \fth(\cythree \times \twoto)
                                &\lolra &\mth(\cythree \times S^1)  
                                &\lolra & \twoa(\cythree)
                                &\lolra &\het(\kthree \times \twoto)  \cr
         \downarrow
                                    &        &\downarrow 
                                    &        &\downarrow
                                    &        &\downarrow \cr
         \fth(\cyfour \times \twoto)
                               &\lolra &\mth(\cyfour \times S^1) 
                               &\lolra & \twoa(\cyfour)
                                &\lolra &\het(\cythree \times \twoto)  \cr
    }
$$
Consider the first line of the diagram, describing dual pairs 
in six dimensions.  We have learned that the gauge group on 
the IIA side is determined by the Picard lattice of the K3. Suppose
we apply the twist map to get a threefold with $k_0=1$, so that
the structure of the gauge group descends from 6d. We have seen that
the threefold contains a singular curve which induces a branching
in the resolution diagram. To push the duality further down,
we apply the twist map to this threefold. 
We see from the structure of the twist map (\ref{twistn}) for 
$n=3$ that for the subclass of fourfolds with $k_0=1$ the orbifolding
essentially embeds the singular curve $C$ of the threefold fiber 
into a singular surface on the fourfold. This leads to a prediction 
for the rank of the group H$^2$ of the fourfold because from our 
previous discussion of the rank of the gauge groups of the models 
dual to IIA(CY$_3$) we expect the rank of H$^2$(CY$_4$) to be 
determined by the rank of these gauge groups \cite{bs96}. In order 
to perform this test of the dualities we need to compute the Hodge 
numbers of the fourfolds.

{\bf Example I:}
We consider an example of the type (\ref{nestfib}) for which 
we choose the fiber to be the threefold $\IP_{(1,1,12,28,42)}[84]$ 
of our first example in the previous Section. The threefold 
is a K3-fibration with an elliptic K3 surface. The twist map 
now simply embeds the singular curve $C=\IP_{(6,14,21)}[42]$ on 
the threefold into the $\ZZ_2$-singular K3 surface 
$\IP_{(1,6,14,21)}[42]$ of the resulting fourfold 
$\IP_{(1,1,2,24,56,84)}[168]$. 
Therefore we expect the gauge structure of the fourfold 
to be determined completely by the threefold and we expect  
$$
\ags ({\rm CY}_4) = \ags ({\rm CY}_3) + 1.
$$
Computing the Hodge diamond \cite{bs96}
\begin{footnotesize}
\beq
\matrix{      &   &       &     &1       &     &      &    &   \cr
              &   &       &0    &        &0    &      &    &   \cr
              &   &0      &     &12       &     &0      &    &   \cr
              &0  &       &0    &        &0    &       &0   &   \cr
          1   &   &27548    &     &110284    &     &27548    &    &1  \cr
      }
\eeq
\end{footnotesize}

\noindent
confirms this expectation. 

{\bf Example II:} 
 For fourfolds a single fiber with different automorphisms 
can lead to topologically distinct fibrations, similar to
the case of fibered threefolds. To illustrate this 
consider the Calabi-Yau threefold configuration  
$\IP _{(1,1,3,3,4)}[12] $ with spectrum $(\ags , \gens) = (5,89)$. 
In this configuration we pick the Fermat hypersurface 
and first consider the image of the
twist map obtained from the automorphism
$\ZZ _{12} : (y_0, y_1, y_2, y_3, y_4) \mapsto (\alpha y_0,
y_1, y_2, y_3, y_4) $, where $\alpha $ is a $12^{th}$ root of
unity. The resulting fourfold lives in the configuration 
$\IP _{(1,1,2,6,6,8)}[24]$ and has the 
Hodge (half-) diamond

\begin{footnotesize}
\beq
\matrix{      &   &       &     &1       &     &      &    &   \cr
              &   &       &0    &        &0    &      &    &   \cr
              &   &0      &     &6       &     &0      &    &   \cr
              &0  &       &0    &        &0    &       &0   &   \cr
          1   &   &722    &     &2956    &     &722    &    &1.  \cr
      }
\eeq
\end{footnotesize}

This fourfold contains the  usual singular 
$\ZZ_2$-surface $\IP _{(1,3,3,4)}$,
whose resolution gives one $(1,1)$-form. On this surface there
are four $\ZZ_4$-points in $\IP_{(1,1)}[4] $ introducing four
further $(1,1)$-forms. Together with the $(1,1)$-form from the
ambient space we get $\ags = 6$.

An alternative possibility to obtain a fourfold with the fiber 
$\IP_{(3,1,1,3,4)}[12] $ is to use the automorphism 
$ \ZZ_{4}: (y_0, y_1, y_2, y_3, y_4) \mapsto (\alpha y_0, y_1,  y_2,
y_3, y_4 ) $, with $\alpha $ a $ 4^{th} $ root of unity. 
Twisting with the appropriate curve $ \cC_{4} $ leads to the
manifold $\IP _{(3,3,2,2,6,8)}[24] $ with Hodge diamond

\begin{footnotesize}
\beq 
\matrix{      &   &       &     &1       &     &      &    &   \cr
              &   &       &0    &        &0    &      &    &   \cr
              &   &0      &     &3       &     &0      &    &   \cr
              &0  &       &9    &        &9    &       &0   &   \cr
          1   &   &254    &     &1054    &     &254    &    &1.  \cr
      }
\eeq
\end{footnotesize}

Again we have a singular $\ZZ_2$-surface $\IP_{(1,1,3,4)}$,
introducing a $(1,1)$-form. Furthermore there is a $\ZZ_3$
curve $\IP_{(1,1,2)}[8]$ introducing the $(2,1)$-forms and
a $(1,1)$-form. 

{\bf Example III:}
In our final example we illustrate the complexity of the 
singularity structure of the fourfolds which can occur 
in configurations with $k_0>1$, depending on the choice 
of automorphism which is being used to construct the 
fibered fourfold.
We start with the threefold $\IP _{(1,2,3,6,6)}[18]$, with 
$(\ags , \gens) =(7,79) $.
Here we have a singular $\ZZ_3 $-curve $\IP_{(1,2,2)}[6] $ and a singular 
$\ZZ_2 $ curve 
$\IP _{(1,3,3)}[9] $ which intersect in the three points $\IP_1[3]$.
A fibration with
this threefold fiber is $\IP_{(1,1,4,6,12,12)}[36]$ with Hodge 
numbers $(h^{(1,1)}=8,h^{(2,1)}=3,h^{(3,1)}=899,h^{(2,2)}=3666)$.

We have the usual $\ZZ_2$-surface on top of which we find the
threefold singularities.
One other possibility to obtain a fibration with this threefold
is to consider $\IP_{(3,3,2,4,12,12)}$ with
Hodge numbers 
$(h^{(1,1)}=8,h^{(2,1)}=1,h^{(3,1)}=321,h^{(2,2)}=358)$.
Here again we have a $\ZZ_2$-surface with a $\ZZ_4$ curve, but 
because this time $k_0 =3 $ we have an additional $\ZZ_3$ surface
instead of a curve. This time the fact that $k_0 > 1 $ has lead
to a higher dimensional singular set because of the particular
form of the weights.

From these results we see \cite{bs96} that the prediction via the 
twist map for the $h^{(1,1)}$-cohomology of the Calabi-Yau fourfolds 
from the dual heterotic gauge groups are confirmed. 

In the next Section we will further generalize the twist map to 
complete intersection spaces of higher codimension. Such manifolds 
  occur in the process of connecting the moduli spaces of 
  different Calabi-Yau manifolds via the splitting type 
  transition in arbitrary complex dimensions. 
  First however we will describe these transitions and
  the resulting unification of vacua.

\section{Unification of vacua}

\subsection{Connecting threefolds} 
A longstanding problem in string theory is the issue of the vacuum 
degeneracy. Soon after the discovery of anomaly free low energy field 
theories of various types of strings the existence of a plethora 
of consistent ground states was shown to exist 
among different compactification 
schemes and 4D string constructions. This appears to create a difficulty 
which in the early days of the first string revival led to some 
disenchantment. Namely, if all these consistent vacua are disjoint then 
this raises the question whether even in principle the string could ever 
determine its own ground state and eventually make some detailed 
predictions. If on the other hand vacua with different spectra 
are connected then this implies that a singularity must occur 
somewhere. It is then not a priori clear that the string can 
consistently propagate in the background of the singular configuration. 
 Thus one has to face the issues of connecting different vacua 
 and providing a physical interpretation of the resulting 
singularities.
Both of these problems admit a solution in the context 
of Calabi-Yau compactifications.

The first of these problems was solved in \cite{cdls88} by 
showing that complete intersection Calabi-Yau 
manifolds with different moduli spaces can indeed be connected. 
In the simplest instance the Calabi-Yau spaces degenerate at the
transition locus at a number of nodes, leading to conifold 
configurations. The resulting conifold transitions have been 
shown to connect all complete intersection Calabi-Yau 
manifolds \cite{cgh90} and have also been shown to generalize 
to the class of  weighted complete intersection spaces \cite{ls95}. 
In the simplest class of transitions hypersurfaces in  
weighted $\IP_4$ \cite{cls90,allg} are connected with codimension two 
varieties. Other possible ways of connecting different manifolds 
involve the extremal transitions of the toric 
framework \cite{mv96, extremal3}. 

At the time when conifold transitions \cite{cdls88} were introduced 
it was unclear what the correct physical 
interpretation is of the divergences associated to the 
conifold transition \cite{cgh90}. Only recently has it been understood 
through the work of Greene, Morrison and Strominger \cite{coniII} in 
type II compactifications, following similar ideas in 
Seiberg-Witten theory, that the 
divergences in the low energy effective action arise because of 
new massless states which are generated at the conifold configuration 
when the volume of vanishing cycles degenerates.   
This then shows that at least in type II string theory the conifold 
transitions of \cite{cdls88} admit a physically reasonable interpretation.
 
\subsection{Making fourfolds, a mirror symmetry test for dualities}
The same problem arises in even more pronounced form in   F-theory 
and M-theory. One consistent compactification type of both of these 
theories involves Calabi-Yau fourfolds, of which there are many more 
than there are threefolds. 
The class of Fermat type hypersurface threefolds embedded in 
weighted $\IP_4$,  for instance, consists of only 147 configurations,  
whereas the number of Fermat hypersurface fourfolds is 3462, 
more than an order of magnitude larger.
Preliminary results show that the number of all weighted hypersurfaces 
consists of at least several hundred thousand 
configurations \cite{lsw97}. This suggests that there are millions 
of complete intersection Calabi-Yau fourfolds, providing possible 
vacua for 4D F-theory.  

A complete construction of this class is not known at present 
and even the enumeration along the lines of \cite{cls90, allg} 
of the subset of all hypersurfaces embedded in weighted 
$\IP_5$ has not been achieved at this point. Some information can 
however be derived by indirect means. 
Via the dualities in Section 2, for instance, 
 we are lead to expect mirror symmetry 
for the space of Calabi-Yau fourfolds because of its relation to 
(0,2) vacua. For the space of (0,2) ground states of the heterotic 
string recent results have established mirror symmetry for a large set 
of Landau-Ginzburg type theories \cite{bsw96}. 
Turning this observation around we see that the  
same dualities support the expectation that the space of heterotic 
(0,2) vacua is indeed vastly larger than the number of (2,2) symmetric 
ground states and that there are many classes of (0,2) theories that 
remain to be discovered (see \cite{bsw96} and references therein).
 
The cohomology of Calabi-Yau threefolds is characterized by only 
two distinct Hodge numbers, leading to the mirror plot 
\cite{cls90,allg} of the combinations 
$(h^{(1,1)}+h^{(2,1)},\frac{\chi}{2}=h^{(1,1)}-h^{(2,1)})$.
For fourfolds the cohomology leads to four independent Hodge numbers 
$(h^{(1,1)},h^{(2,1)},h^{(3,1)},h^{(2,2)})$ with the duality 
relations $h^{(p,q)}=h^{(4-p,4-q)}=h^{(q,p)}$. A proper Hodge type 
plot thus would be four-dimensional, somewhat difficult to visualize. 

Mirror symmetry among Calabi-Yau fourfolds however distinguishes 
again the combinations 
$(h^{(3,1)}-h^{(1,1)}$, $h^{(3,1)}+h^{(1,1)}$) because both 
$h^{(2,1)}$ and $h^{(2,2)}$ remain invariant under the mirror 
flip $h^{(p,q)}(M) = h^{(D-p,q)}(M')$ of a mirror pair $(M,M')$. 
The standard mirror plot \cite{cls90} therefore is a measure 
of mirror symmetry in the present class as well.  
In Figure 1 we have plotted these combinations within a lower 
range of these variables, cutting off the Hodge numbers at about 
40k. This restriction cuts off those hypersurfaces whose degree is 
in the range of millions for which the Hodge numbers can easily 
reach the x00$k$ range.  
 
\par\noindent
  \centerline{\epsfxsize=4.5in\epsfbox{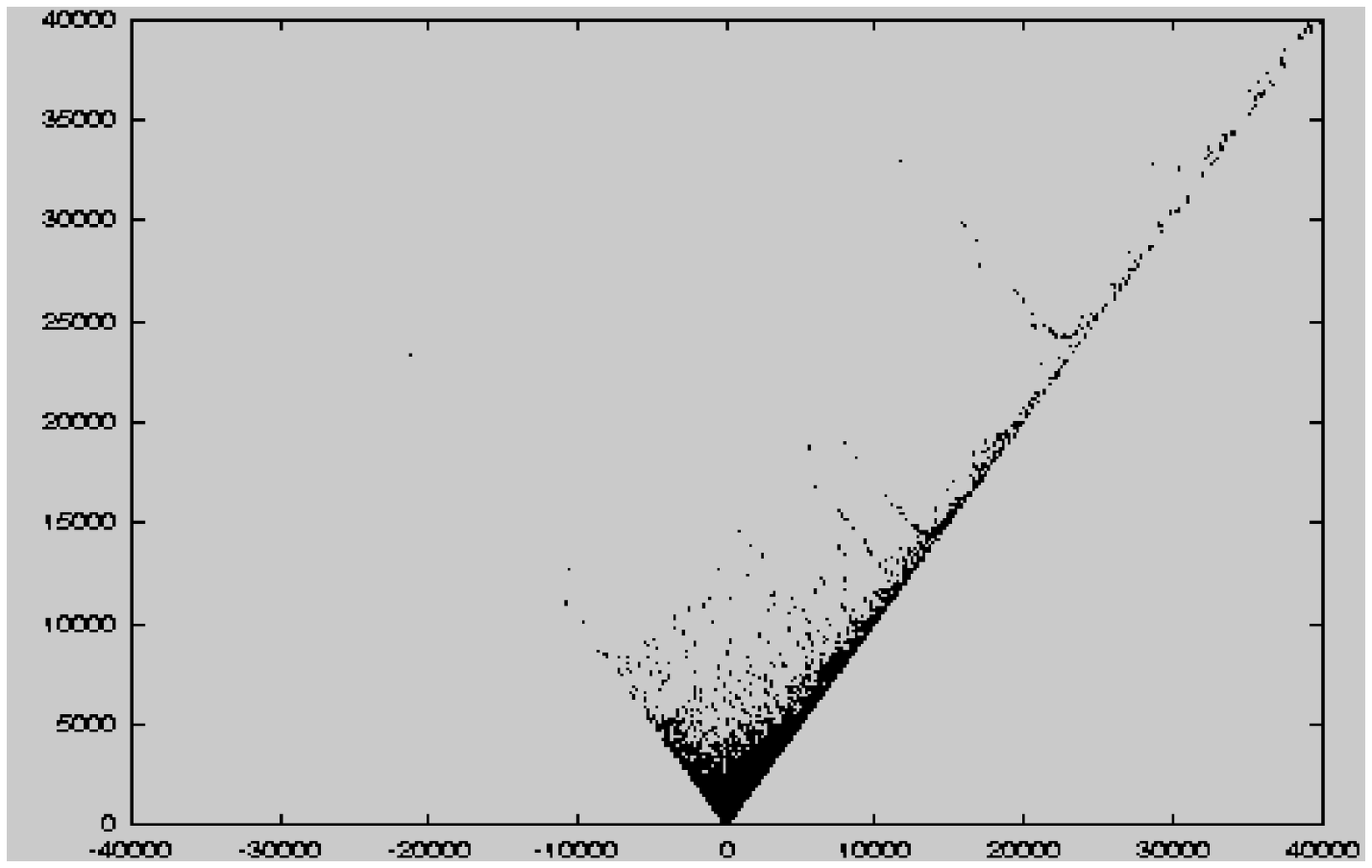}}
\par\noindent

\noindent
{\bf Figure 1:}{\it ~Plot of $(h^{(3,1)}-h^{(1,1)})$ vs. 
                     $h^{(3,1)}+h^{(1,1)})$ for some tens of 
        thousands of Calabi-Yau fourfolds.}

A zoom-in of Figure 1 into a smaller range of Hodge numbers with 
a cut-off for the Hodge numbers at about 1k reveals 
the expected mirror symmetric structure, as shown in Figure 2. 

\par\noindent
  \centerline{\epsfxsize=4.5in\epsfbox{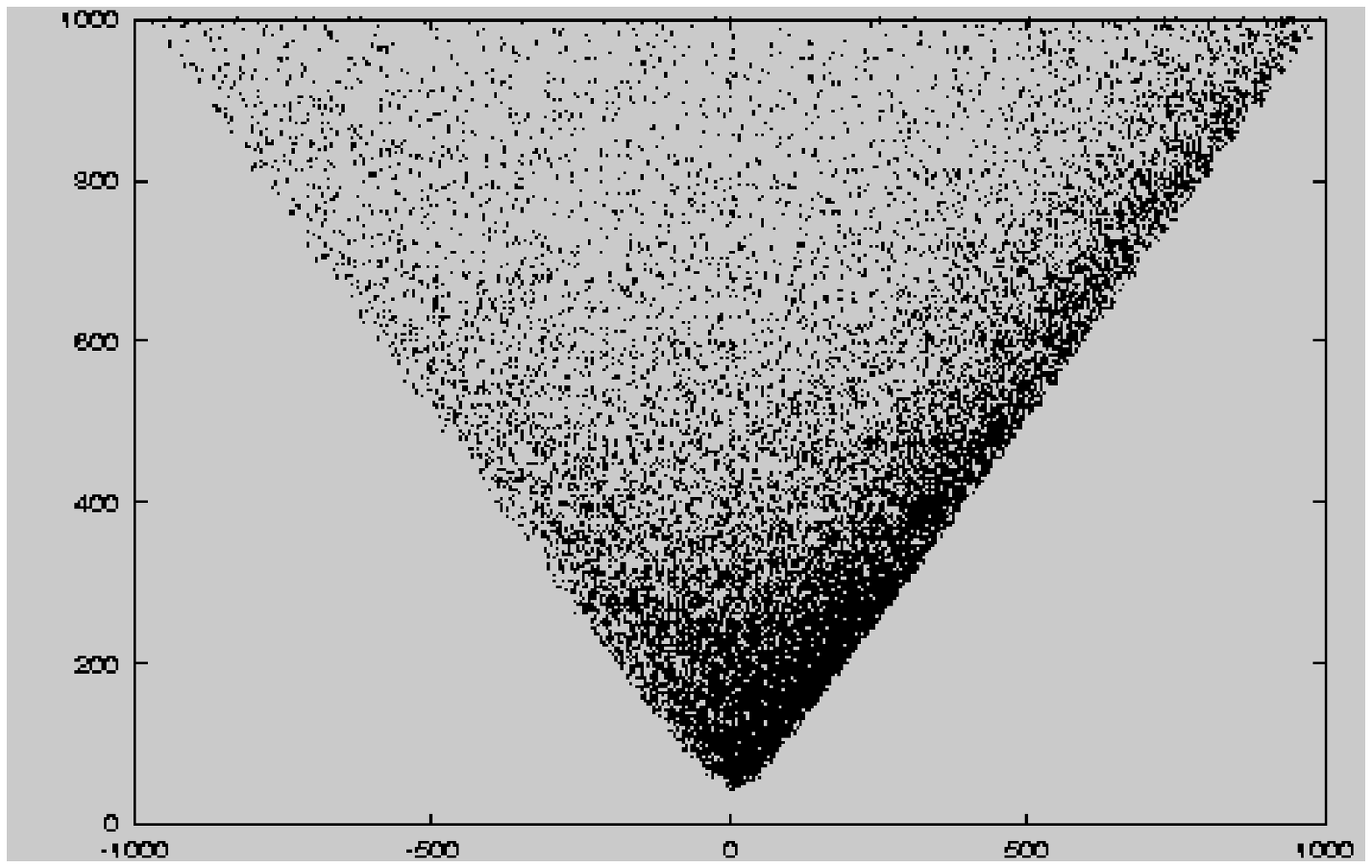}}
\par\noindent

\noindent
{\bf Figure 2:}{\it ~The CY$_4$ plot for a smaller region of 
              Hodge numbers.}

Comparing Figs. 1 and 2 suggests that the construction still has 
to go some distance. Results in this direction have also been 
obtained in refs. \cite{klry97,ks97}.

 As mentioned above we expect the moduli space of CY$_3$-fibered 
fourfolds to be connected via degenerations of the fibers over 
the base space. By using the twist map described in Section 
2 it is possible to construct such transitions 
explicity \cite{bls96}. We will see however that the connectedness is in 
fact more general property, independent of any fibration properties. 
In the present Section we will discuss a particular type of transition 
generalizing the splitting type transition to fourfolds. Whereas 
in the case of threefolds this transition leads to varieties which 
degenerate at a configuration of nodes in the case of fourfolds 
the splitting type transition leads to more severe singular varieties 
which degenerate at a configuration of singular curves \cite{bls96}. 
Extremal types of transitions have been generalized to fourfolds 
in refs. \cite{pm96,klry97}.

\subsection{The splitting transition between weighted complete 
        intersections}

In arbitrary dimensions it is natural to consider 
configurations of the type
\beq
\matrix{\IP_{(k_1^1,\dots ,k_{n_1+1}^1)}\cr
               \IP_{(k_1^2,\dots ,k_{n_2+1}^2)}\cr \vdots \cr
               \IP_{(k_1^F,\dots ,k_{n_F+1}^F)}\cr}
\left [\matrix{d_1^1&d^1_2&\ldots &d_N^1\cr
               d_1^2&d_2^2&\ldots &d_N^2\cr
               \vdots&\vdots&\ddots &\vdots\cr
               d_1^F&d_2^F&\ldots &d_N^F\cr}\right]~~=~~X.
\lleq{wcicys}
Such configurations describe the intersection of the zero locus of $N$
 polynomials embedded in a product of weighted projective spaces,
where $N= \left(\sum_{i=1}^F n_i -D\right)$ is the number of 
polynomials $p_a$ of F-degree $(d_a^1,\dots,d_a^F)$ and
$D$ is the dimension of the manifold described by this degree matrix.
Even though our considerations can be applied to general intersection 
spaces our main interest is in manifolds for which the first Chern class 
\beq\label{chernone}
c_1(X) = \sum_{i=1}^F\left[\sum_{l=1}^{n_i+1} k_l^i -
       \sum_{a=1}^N d_a^i \right]h_i  
\eeq
vanishes. Here we denote by $h_i, i=1,...,F$ the pullback of the generators 
of $H^2(\IP_{(k_1^i,\dots ,k_{n_1+1}^i)})$. 
For simplicity we will first write down a split between intersections
in ordinary projective spaces. 

Introducing two vectors $u,v$ such that $$(u^i+v^i)=d_1^i$$
and denoting the remaining $(F \times (N-1))$-matrix by $M$, we 
write the manifolds (\ref{wcicys}) 
as $Y[(u+v)~~M]$. The simplest kind of transition 
is the $\IP_1$-split which is defined by  
\beq\label{ponesplit}
X~=~Y[(u+v)~~M] ~\lolra ~\matrix{\IP_1\cr Y\hfill \cr} 
\left[\matrix{1&1&0\cr u&v&M\cr}\right]~=~X_{\rm split}. 
\eeq
The split variety of the rhs is described by the polynomials 
of the original manifold and two additional polynomials,  
which we can write as 
\bea \label{splitpolly}
p_1 &=&x_1Q(y_i) + x_2R(y_i)  \nn \\
p_2 &=&x_1S(y_i) + x_2T(y_i),  
\eea
where $Q(y_i), R(y_i)$ are of multi-degree $u$ and $S(y_i), T(y_i)$ 
are of degree $v$. 
In (\ref{splitpolly}) we collectively denote the coordinates of the space 
$Y$ by $y_i$ whereas the $x_i$ are the coordinates of the projective 
line $\IP_1$. To understand the relation between the two
manifolds in (\ref{ponesplit}) we consider (\ref{splitpolly}) in more 
detail. We can regard the vanishing locus of (\ref{splitpolly}) as a 
linear equation system in $x_1$ and $x_2$. These equations only have 
nontrivial solutions if the determinant
$$
p_{det} = QT - RS
$$
vanishes. Together with the original polynomials the determinant 
defines a determinantal variety $X^{\sharp} $. The polynomial $p_{det}$ 
is of multidegree $u+v$ in the
variables of $Y$ and therefore $ X^{\sharp} \in Y[(u+v)~~M]$ where 
\beq\label{detervar}
X^{\sharp} =\{~p_{\rm det} = QT - RS=0,~p_a=0,~a=2,...,N\} 
\eeq
The space $X^{\sharp}$ however is not a smooth manifold. It is
singular on the locus where the determinant has a double zero
because the polynomials $Q,R,S,T $ vanish simultaneously. The
singular set is described by
\beq\label{singloc}
\Si = Y[u~~u~~v~~v~~M]. 
\eeq
The manifold on the lhs in (\ref{ponesplit}) is $D$-dimensional, 
therefore $\Si $ has dimension $(D-3)$. 

\subsection{Splitting and contracting Calabi-Yau fourfolds} 

In the following we will focus on varieties of complex dimension 
three and four. For a threefold split (\ref{singloc})
describes a number of points, i.e. a conifold configuration,
whereas for a fourfold split the singular set is an algebraic curve, 
i.e. a real two-dimensional surface, with in general several 
components. Finally, to move from the rhs of (\ref{ponesplit}) 
to the lhs we have to smooth out the singularity. This can be done 
by adding a transverse piece to the nontransverse polynomial 
$p_{det} $
\beq
p_{\rm def}=p_{\rm det} + t\cdot p_{\rm trans}.  
\eeq
The process to start from the rhs of (\ref{ponesplit}) and to come
to the lhs in the way described above is referred to as
contraction \cite{cdls88}.
The important point however is that the singular set also
admits a small resolution which takes us via the singular variety
$X^{\sharp}$ from the lhs
to the rhs. This means that the singular set is smoothed out
using an exceptional set of codimension two. 
For threefolds this involves the projective line $\IP_1$ and
for fourfolds the projective plane $\IP_2$. Performing such a 
small resolution leads to the higher codimension split manifold.
To support this picture a relation between the Euler numbers of
$X$ and $X^{\sharp}$ was proven in \cite{cdls88} for threefolds
and in \cite{bls96} for fourfolds. 
For threefolds the result is
(neglecting the phenomenon of colliding singularities for 
the moment)
\beq \label{teurel}
\chi (X_{split}) = \chi (X) + 2n,
\eeq
where $n$ denotes the number of singular points
and the factor 2 originates from the Euler number of the $\IP_1$ 
used to smooth out the manifold. 
The analogous formula for fourfolds 
\beq\label{eulrel}
\chi (X_{\rm split}) = \chi (X) + 3 \chi (\Si )
\eeq
describes the resolution involving a $\IP _2 $ whose Euler number
is three.

Thus we arrive at the same singular space by degenerating two 
distinct manifolds in different ways 
$$ X \lra X^{\sharp} \longleftarrow X_{\rm split}.  
$$
Put differently, we can start from a determinantal variety and 
smooth out the singularities in two distinct ways 
$$ X \longleftarrow X^{\sharp} \longrightarrow X_{\rm split}.$$ 
A simple example for a split between two threefolds is
\beq \label{quintsplit}
\IP_4[5]_{-200} ~~\lolra ~~
 \matrix{\IP_1\cr \IP_4\cr} 
     \left[\matrix{1&1\cr 1&4\cr}\right]_{-168}.
\eeq
This conifold transition connects the quintic in $\IP_4 $ with
Hodge numbers $(\ags ,\gens ) = (1, 101) $ to the codimension
two configuration on the rhs with Hodges $(\ags , \gens) = (2,86)$.
The physical interpretation in the context of type II string 
theory of this transition has been discussed in \cite{coniII}. 

The perhaps simplest example of a splitting transition involving fourfolds
is the split of the sextic 
\beq \label{sexsplit}
\IP_5[6]_{2610} ~~\lolra ~~
 \matrix{\IP_1\cr \IP_5\cr}\left[\matrix{1&1\cr 1&5\cr}\right]_{2160}, 
\eeq
where the subscripts denote the Euler numbers.
The smooth hypersurface can be defined by the Fermat polynomial 
$p = \sum_i z_i^6$
and a transverse choice of the split configuration is provided 
by 
\bea \label{sexcodtwo}
p_1 &=& x_1y_1 + x_2y_2 \nn \\  
p_2 &=& x_1\left(y_2^6 + y_4^6 +y_6^6\right) 
        +x_2\left(y_1^6 + y_3^6 + y_5^6\right).\nn 
\eea
The singular set of the determinantal variety 
is given by the genus $g=76$ curve
$\Si=\IP_3[5~~5]$. Thus we can verify the relation (\ref{eulrel}).
More precisely the split (\ref{sexsplit}) connects
the Hodge diamond of the sextic hypersurface 

\begin{footnotesize}
\beq
\matrix{      &   &       &     &1       &     &      &    &   \cr
              &   &       &0    &        &0    &      &    &   \cr
              &   &0      &     &1       &     &0      &    &   \cr
              &0  &       &0    &        &0    &       &0   &   \cr
          1   &   &426    &     &1752    &     &426    &    &1.  \cr
      }
\eeq
\end{footnotesize}
\noindent 
with the Hodge diamond
\begin{footnotesize} 
\beq\label{mhodge1}
\matrix{      &   &       &     &1       &     &      &    &   \cr
              &   &       &0    &        &0    &      &    &   \cr
              &   &0      &     &2       &     &0      &    &   \cr
              &0  &       &0    &        &0    &       &0   &   \cr
          1   &   &350    &     &1452    &     &350    &    &1.  \cr
      }
\eeq
\end{footnotesize}
of the codimension two complete intersection manifold  of 
(\ref{sexsplit}).

It is of interest to generalize splitting for three- and four-
dimensional manifolds to complete intersections in weighted
projective spaces. In the weighted context it will also be
possible to apply the twist map, which helps to investigate  
transitions between fibered manifolds. 

\subsection{Weighted transitions between K3-fibered threefolds}
  
The weighted conifold transitions between threefolds of
\cite{ls95} are of the general form:
\beq\label{threesplit}
\IP_{(k_1,k_1,k_2,k_3,k_4)}[d]~\lolra ~
\matrix{\IP_{(1,1)} \hfill \cr \IP_{(k_1,k_1,k_2,k_3,k_4)}\cr} 
\left[\matrix{1&1\cr k_1&(d-k_1)\cr}\right]
\eeq
with $d=2k_1+ k_2+ k_3+ k_4$. 

In general conifold transitions connect K3-fibered manifolds 
with spaces which are not fibrations. An example 
is the transition from the quasismooth octic $\IP_{(1,1,2,2,2)}[8]$   
to the quintic $\IP_4[5]$. It can be shown, however,  
that certain types of weighted conifold transitions 
exist which do connect K3 fibered manifolds.

A simple class of such conifold transitions  between 
fibered manifolds is provided by the weighted splits 
summarized in the diagram \cite{ls95}  
\beq
\IP_{(2l,2l,2m,2k-1,2k-1)}[2(d+l)]~\lolra ~
   \matrix{\IP_{(1,1)}\hfill \cr \IP_{(2l,2l,2m,2k-1,2k-1)}\cr}
       \left[\matrix{1&1\cr 2l&2d\cr}\right],    
\eeq
where $d=(2k-1+l+m)$. 
Here the hypersurfaces, containing the K3 surfaces
$\IP_{(2k-1,l,l,m)}[2k-1+2l+m]$, split into codimension two manifolds
which contain the K3 manifolds
\beq
\matrix{\IP_{(1,1)}\hfill \cr \IP_{(l,l,m,2k-1)}\cr}
\left[\matrix{1&1\cr l&d\cr}\right]
\lleq{splitk3}
of codimension two.

\subsection{Twist map for split manifolds}

In order to see the detailed fiber structure of the above 
varieties of codimension two it is useful to generalize the 
twist map we described in Section 2 to 
complete intersection manifolds. 
Consider the K3 surfaces of the type (\ref{splitk3}) and the 
associated curves $\IP_{(2,1,1)}[2d]$ with $d=(l+m+2k-1)$. 
With these ingredients 
we can define a generalized twist map 
\beq 
\IP_{(2,1,1)}[2d] \times 
  \matrix{\IP_{(1,1)}\hfill \cr \IP_{(l,l,m,2k-1)}\cr}
  \left[\matrix{1&1\cr l&d\cr}\right]
~\lra ~
 \matrix{\IP_{(1,1)}\hfill \cr \IP_{(2l,2l,2k-1,2k-1,2m)}\cr}
       \left[\matrix{1&1\cr 2l&2d\cr}\right] 
\eeq
via 
\beq
\left((x_0,x_1,x_2),(u_0,u_1),(y_0,...,y_3)\right) 
     ~\lra ~\left((u_0,u_1),
         (x_1\sqrt{\frac{y_0}{x_0}},x_2\sqrt{\frac{y_0}{x_0}},
              y_1,y_2,y_3)\right).
\eeq
Again we see that the quotienting introduces additional singular 
sets and the remarks of the previous Section apply in the present 
context as well. In particular we see the new singular curve 
\beq 
\ZZ_2:~~C=\matrix{\IP_{(1,1)}\hfill \cr \IP_{(l,l,m)}\cr}
  \left[\matrix{1&1\cr l&d\cr}\right]
\eeq
which emerges on the threefold image of the twist map.

An example for such a transition between manifolds which are
K3-fibered as well as elliptically fibered is given by
\beq
\IP_{(1,1,2,4,4)}[12] ~\lolra ~
\matrix{\IP_{(1,1)}\hfill \cr \IP_{(4,4,1,1,2)}\cr}
\left[\matrix{1&1\cr 4&8\cr}\right] \nn \\   
\lleq{hetsplit}
where the Hodge numbers of the hypersurface are  
$(h^{(1,1)},h^{(2,1)})=(5,101)$ while those of the 
codimension two threefold $(h^{(1,1)},h^{(2,1)})=(6,70)$.  
Here the transverse codimension two variety of the rhs 
configuration is chosen to be 
\bea
p_1 &=& x_1y_1 + x_2y_2 \nn \\ 
p_2 &=& x_1(y_2^2+y_4^8+y_5^4) + x_2(y_1^2+y_3^8-y_5^4) 
\eea
which leads to the determinantal variety in the lhs 
hypersurface configuration 
\beq
p_{\rm det}=y_1^3 -y_2^3 +(y_1y_3^8 
      - y_2y_4^8)-(y_1+y_2)y_5^4.  
\lleq{detvar}
This variety is singular at 
$\IP_{(4,4,1,1,2)}[4~4~8~8] = 32$ nodes, which can be resolved 
by deforming the polynomial. 

Contained in these 2 CY-fibrations are the K3 configurations
\beq
\IP_{(2,2,1,1)}[6] ~\lolra ~
\matrix{\IP_{(1,1)}\hfill \cr \IP_{(2,2,1,1)}\cr}
\left[\matrix{1&1\cr 2&4\cr}\right],
\lleq{k3split2}
where the left-right arrow indicates that these two
are indeed related by splitting. The K3 of the rhs is
obtained by considering the divisor
\beq
D_{\th} = \{y_4=\th y_3\}
\lleq{div1}
which leads to
\beq
\matrix{\IP_{(1,1)}\hfill \cr \IP_{(4,4,1,2)}\cr}
\left[\matrix{1&1\cr 4&8\cr}\right],
\eeq
with
\bea
p_1&=&x_1y_1 + x_2y_2 \nn \\
p_2&=& x_1(y_2^2 + \th^8 y_3^8 + y_5^4)
         + x_2(y_1^2 + y_3^8 - y_5^4). \nn \\
\eea
Because of the weights in the weighted $\IP_3$ this is equivalent
to the rhs of (\ref{k3split2}) with
\bea
p_1 &=&x_1y_1 + x_2y_2 \nn \\
p_2 &=& x_1(y_2^2 + \th^8 y_3^4 + y_5^4)
         + x_2(y_1^2 + y_3^4 - y_5^4). \nn \\
\eea
The determinantal variety following from this space is
\beq
p_s = y_1^3-y_2^3 +(y_1 - \th^8 y_2)y_3^4 - (y_1-y_2)y_5^4.
\eeq
But this is precisely what one gets by considering the divisor
in the determinantal 3-fold variety (\ref{detvar}) and thus
we see that the conifold transitions take place in the fiber
of the CY 3-fold.

\subsection{Weighted fourfold transitions via fiber degenerations}

We will now look at fourfold transitions originating from
threefold transitions via the twist map.
We want to apply the twist map to the transitions (\ref{threesplit})
in order to obtain fibered fourfolds.
Let $\ell=d/k_4 \in 2\IN +1$. For the hypersurfaces of (\ref{threesplit}) this 
amounts to choosing the curve $\cC_{\ell} = \IP_{(2,1,1)}[2\ell]$ and 
applying the twist map 
\beq\label{hypetwist}
\IP_{(2,1,1)}[2\ell] ~\times~ \IP_{(k_1,k_1,k_2,k_3,k_4)}[d]  
 ~\lra ~\IP_{(2k_1,2k_1,2k_2,2k_3,k_4,k_4)}[2d] 
\eeq
defined as 
\beq\label{hypemap}
\left((x_1,x_2,x_3),(y_1,y_2,y_3,y_4,y_5)\right)~~\mapsto ~ 
\left(y_1,y_2,y_3,y_4,x_2\sqrt{y_5\over x_1}, x_3\sqrt{y_5\over x_1}\right).
\eeq

For the codimension two threefold in (\ref{threesplit}) the twist map 
produces the complete intersection fourfolds 

\beq
\IP_{(2,1,1)}[2\ell] ~\times ~
\matrix{\IP_{(1,1)}\hfill \cr \IP_{(k_1,k_1,k_2,k_3,k_4)}\cr}
\left[\matrix{1&1\cr k_1&(d-k_1)\cr}\right]~\lra ~
\matrix{\IP_{(1,1)} \hfill \cr \IP_{(2k_1,2k_1,2k_2,2k_3,k_4,k_4)}\cr}
\left[\matrix{1&1\cr 2k_1&2(d-k_1)\cr}\right]. 
\lleq{codimtwotwist}

\relax From this we see that the twist map applied to threefolds which are 
connected via conifold transitions induces splitting transitions 
between fibered fourfolds 
\beq\label{foursplit}
\IP_{(2k_1,2k_1,2k_2,2k_3,k_4,k_4)}[2d]~\lolra~
\matrix{\IP_{(1,1)} \hfill \cr \IP_{(2k_1,2k_1,2k_2,2k_3,k_4,k_4)}\cr}
\left[\matrix{1&1\cr 2k_1&2(d-k_1)\cr}\right].  
\eeq
In this way the twist map maps the threefold split (\ref{hetsplit}) in
the previous Section to the fourfold split
\beq\label{weightsplit}
\IP_{(8,8,4,2,1,1)}[24]~~\lolra ~ 
\matrix{\IP_{(1,1)}\hfill \cr \IP_{(8,8,4,2,1,1)}\cr}
\left[\matrix{1&1\cr 8&16\cr}\right],\nn   
\eeq
where the lhs manifold is defined by the zero locus of the polynomial  
$$
p=z_0^3+z_1^3+z_2^6+z_3^{12}+z_4^{24}+z_5^{24} 
$$
and the rhs by the equations
\bea
p_1 &=& x_1y_1 + x_2y_2  \nn \\
p_2 &=& x_1(y_2^2 + y_4^8 + y_6^{16})
        + x_2(y_1^2 + y_3^4 + y_5^{16}). \nn
\eea
The determinantal variety determined by 
$$
p_{det} = y_1(y_1^2 + y_3^4 + y_5^{16}) -
     y_2(y_2^2 + y_4^8 + y_6^{16}) 
$$
is singular on the locus $\Si = \IP_{(4,2,1,1)}[16~16]$, describing a 
smooth curve of genus $g=385$.
Altogether, the fibration structure of the lhs of this example
can be summarized by several applications of the twist map as
\bea
\IP_2[3] ~\lra ~\IP_{(2,2,1,1)}[6] ~\lra ~ \IP_{(4,4,2,1,1)}[12] 
   ~\lra ~\IP_{(8,8,4,2,1,1)}[24],
\eea
whereas the codimension two space leads to the iterative structure 

\beq
\matrix{\IP_1\cr \IP_2\cr}\left[\matrix{1&1\cr 1&2\cr}\right]
~\lra ~
\matrix{\IP_{(1,1)}\hfill \cr \IP_{(2,2,1,1)}\cr}
        \left[\matrix{1&1\cr 2&4\cr}\right]
~\lra ~
\matrix{\IP_{(1,1)}\hfill \cr \IP_{(4,4,2,1,1)}\cr}
        \left[\matrix{1&1\cr 4&8\cr}\right]
~\lra ~
\matrix{\IP_{(1,1)}\hfill \cr \IP_{(8,8,4,2,1,1)}\cr}
        \left[\matrix{1&1\cr 8&16\cr}\right]. 
\eeq
Both sides of this split provide examples for the nested fibration
structure (\ref{nestfib}) mentioned in the introduction.

\section{Superpotentials}

In the previous Section we have seen that some regions of M- and
F-theory vacua are connected by transitions. In the present Section
we will see that these vacua can be distinguished in an intrinsic
manner by the property that some of them lead to a nonperturbatively
generated superpotential.
In \cite{wsp96} it was shown that a nonperturbative superpotential
is generated for compactifications of M-theory on certain
Calabi-Yau fourfolds. An example which shows modular behaviour
for the superpotential was described in \cite{dgw96}. In \cite{bls96}
the variety of \cite{dgw96} was connected to a manifold which
does not generate a superpotential by a splitting transition.
Before we describe this transition we will  briefly review 
the generation of superpotentials in three-dimensional
field theory and M-theory compactifications to three dimensions.

\subsection{The superpotential in 3D field theory} \label{fsp}

The generation of a superpotential by instantons
in $d=3$, $N=2$ field theory was considered in \cite{ahw82}
via dimensional reduction of an
$N=1$ supersymmetric gauge theory with gauge group
$SU(2)$ from four dimensions.
It was shown that due to instanton effects
there are potential terms for the scalar $\varphi $
arising in the three-dimensional theory as a mode of the
4-dimensional gauge field.

Recall that there are two types of
supersymmetric invariants in 4 dimensions (and this is also
valid for the 3D gauge theories under consideration).
\bea
I_1 &=& \int d^4x ~d^2\theta ~d^2{\overline{\theta }}~~
       Q(x,\theta , {\overline{\theta }} ) \\
I_2 &=& \int d^4x ~d^2\theta ~~R(x, \theta ),
\eea
where $Q$ and $R$ are superfields. Because $R$ does not depend on
$\overline{\theta} $ we can build a supersymmetric invariant without
integration over $\overline{\theta } $.
The invariants of type $I_1$ contain the kinetic terms, whereas
those of type $I_2$ contain the mass terms and Yukawa couplings.
 There are nonrenormalization theorems saying that 
to any finite order in perturbation theory
quantum corrections can only affect operators which can be written in
the form $I_1$. Therefore we have to take into account
nonperturbative effects if we want to generate a superpotential.
The instantons in the 3D gauge theory are monopoles in the
BPS-limit. They are invariant under half of the four supercharges.
Two supercharges generate two fermion zero modes.
Instanton corrections lead to a factor $e^{-I}$
in the effective action, where $I$ is the one-instanton action.
In the BPS-case it is given by
$$
I=\frac{4\pi \varphi }{e}
$$
In 3D we are in the situation that the gauge field is dual to
a scalar $\phi$. If one computes the effective action in 3D 
it can be seen that the scalar field $\varphi $ combines with
the field $\phi $ to a complex field $ Z=\varphi + i \phi.$
The instanton correction becomes
\beq \label{inst}
e^{-(I+i\phi)}
\eeq
Actually, $\varphi $ and $ \phi $ have combined to give the scalar
component of a chiral superfield.

The effect of the fermion zero modes is that the function
(\ref{inst}) must be integrated over chiral superspace
\beq \label{instpot}
\int d^2 \theta ~~e^{-(I+i\varphi )}
\eeq
and
is a superpotential rather than an ordinary potential.

\subsection{Superpotentials in M-theory compactifications}

Consider now the compactification of M-theory
to three dimensions on a Calabi-Yau fourfold. It was shown 
in \cite{wsp96} that a superpotential can be generated by
wrapping 5-branes over certain divisors in the fourfold, 
resulting in the M-theory analogs of the field theory quantities
reviewed  in the previous Section.

Gauge fields in three dimensions are obtained as modes of
the 3-form potential $C$
\beq
C_{\mu i\bar{j}} ~(x,y) =
\sum_{\Lambda =1}^{\ags } A^{\L}_\mu (x) ~
   \omega_{i\bar{j}}^{(\Lambda )}.
\eeq
Here, $\mu $ is the three-dimensional spacetime index and $i,j$ are
internal indices. $ \omega_{i \bar{j}}^{(\Lambda )} $ are a
basis of (1,1)-forms of the fourfold. Again, the $A^{\L}_\mu$'s are 
dual to scalars. The next thing we have to look for is the analog
of the BPS-monopoles of the field theory.
The magnetic source for $C$ is the
M-theory M5-brane. Thus, we can produce something that looks
like an instanton in 3D by wrapping the world-volume of the
M5-brane around 6-cycles in the Calabi-Yau. As we have seen in
the previous Section we  need the property
that the instanton must be invariant under two of the four supersymmetries,
so the cycle must be a complex divisor.
The field theory analysis of the previous Section leads us to
expect a superpotential of the form
\beq \label{mpot}
e^{-(V_D + i\phi _D)}~P,
\eeq
where $V_D$ is the instanton-action determined by the volume of
the divisor $D$ around which the M5-brane wraps, and $\phi _D$ 
is a linear combination of dual scalars.
$P$ is a one-loop determinant of world-volume fields, whose
zeroes determine the vacua of the theory. In \cite{wsp96} the
fermion zero modes were discussed. Using an anomaly-cancellation
argument it  was derived that the divisor
has to fulfill additional properties to give a nonvanishing
contribution to the superpotential.
A particularly simple case is the situation where the only 
fermionic zero modes are the two fermion zero modes coming from 
the supercharges. Here a superpotential is generated
(see Section \ref{fsp}). 
One way to guarantee that there are no 
further fermion zero modes is to require that the divisor has 
to satisfy the conditions 
\beq \label{cond}
\dim \Omega^{(0,1)} = \dim \Omega^{(0,2)} 
 = \dim \Omega^{(0,3)} = 0.
\eeq
The reason for this is that on a K\"ahler manifold the
zero modes of the Dirac operator are given by the Dolbeault-
cohomology.  We then obtain a superpotential similar to 
(\ref{instpot}).

The condition (\ref{cond}) implies that the arithmetic
genus of the divisor equals one.
\beq \label{gencond}
\chi (D, \cO_D )= \sum_{n=0}^{3} (-1)^n ~\dim \Omega^{(0,n)} = 1.
\eeq
The precise anomaly cancellation argument 
shows that (\ref{gencond}) has to hold also in more general
situations where a superpotential is generated, i.e. (\ref{gencond})
is a necessary (but not sufficient) condition for superpotential
generation, whereas (\ref{cond}) is sufficient but not necessary.

\subsection{Generating a superpotential via splitting}
\noindent
We will now consider the behaviour of the superpotential
under splitting transitions. Consider the manifold
\beq \label{dgwcontract}
X=\matrix{\IP_1\cr \IP_2\cr \IP_2\cr}
      \left[\matrix{2\cr 3\cr 3\cr}\right].
\eeq
\relax From Lefschetz' hyperplane theorem
we know that $h^{(1,1)}=3$ and $h^{(2,1)}=0$. Furthermore we
can determine $h^{(3,1)}=280$ by counting complex deformations.
Plugging all this into the Euler number leads to the complete
Hodge half-diamond

\begin{footnotesize}
\beq \label{mhodge2}
\matrix{      &   &       &     &1       &     &      &    &   \cr
              &   &       &0    &        &0    &      &    &   \cr
              &   &0      &     &3       &     &0      &    &   \cr
              &0  &       &0    &        &0    &       &0   &   \cr
          1   &   &280    &     &1176    &     &280    &    &1.  \cr
      }
\eeq
\end{footnotesize}

It has been shown in \cite{wsp96}
 that manifolds of this type, i.e. hypersurfaces
embedded in products of ordinary projective spaces, do not lead to
nonvanishing superpotential. However the manifold above
 can be split into one that does contain divisors which generate
a superpotential, namely the manifold studied in \cite{dgw96}.
\beq
\matrix{\IP_1\cr \IP_2\cr \IP_2\cr}
\left[\matrix{2\cr 3\cr 3\cr}\right]
~\lolra ~~
\matrix{\IP_1\cr \IP_1\cr \IP_2\cr \IP_2\cr}
\left[\matrix{1&1\cr 2&0\cr 3&0\cr 0&3\cr}\right] = X_{\rm split}.
\eeq
Both of these spaces are elliptic fibrations and the split manifold is
also a K3-fibration with generic elliptic K3 fibers.

The determinantal hypersurface
\beq
X^{\sharp} = \{p_{\rm det} = QT -RS=0\} 
       \in \matrix{\IP_1\cr \IP_2\cr \IP_2\cr}
\left[\matrix{2\cr 3\cr 3\cr}\right]  
\eeq
is singular at the locus
\beq
\matrix{\IP_1\cr \IP_2\cr \IP_2\cr}
\left[\matrix{2&2&0&0\cr 3&3&0&0\cr 0&0&3&3\cr}\right]= 9\times \Si,
\eeq
where $\Si = \matrix{\IP_1\cr \IP_2\cr}
\left[\matrix{2&2\cr 3&3\cr}\right] $
and $\IP_2[3~~3]=9pts$. The curve
$\Si$ has Euler number $\chi(\Si) =-54$ and hence genus
$g(\Si)=28$. Thus the singular set has 9 different components and
the splitting formula (\ref{eulrel}) becomes
\beq
\chi(X_{\rm split}) = \chi(X) + 3\cdot 9~\chi(\Si)=288.
\eeq
We see from this that it is precisely the small resolution of the
curve $\Si$ which introduces the divisors in $X_{\rm split}$ which are
responsible for the superpotential. But splitting does not necessarily
change the superpotential. Starting from a manifold with vanishing
superpotential there can also be splits which connect this manifold
with another manifold with no superpotential. An example for this is
the sextic split (\ref{sexsplit}).

\vskip .2truein 
\noindent
{\bf Acknowledgement}

\noindent
R.S. thanks  the Theory Group at UT Austin for hospitality.
It is a pleasure to thank the members of the Theory Group and 
the Math/Phys String Seminar for discussions.  
This work was supported in part by NSF grant PHY-95-11632.

\end{document}